\documentclass[aps,prl,twocolumn,superscriptaddress,amsmath,amsfonts,showpacs]{revtex4}

\usepackage{graphicx}

\begin{document}


\title{
Hall Effect of Spin Waves in Frustrated Magnets
}


\author{Satoshi Fujimoto}
\affiliation{Department of Physics, Kyoto University, Kyoto 606-8502, Japan}




\date{\today}

\begin{abstract}
We examine a possible spin Hall effect for localized spin systems
with no charge degrees of freedom.
In this scenario, a longitudinal magnetic field gradient induces
a transverse spin current carried by 
spin wave excitations with
an anomalous velocity which is caused by a topological Berry-phase effect 
associated with spin chirality,
in analogy with anomalous Hall effects 
in itinerant electron systems.
Our argument is based on a semiclassical equations of motion 
applicable to general spin systems.
Also, a microscopic model of frustrated magnets which exhibits
the anomalous spin Hall effect is presented.
\end{abstract}



\maketitle



{\it Introduction ---}
Spin transport phenomena in condensed matter systems
have been attracting much interest
because of potential applications to spintronics~\cite{dya,kato},
and also because of their fundamental relation with
the notion of
topologically-induced spin currents~\cite{niu,niu2,juig}.
For semiconductors and metals with spin-orbit interactions,
the spin Hall effect (SHE) in the absence of magnetic fields
was predicted theoretically~\cite{dya}, 
and later experimentally realized~\cite{kato}. 
This effect allows for topological interpretation in terms of
the Berry phase;
the spin-orbit interaction gives rise to the Berry curvature
associated with the Bloch wave function, which results in  
the nonzero anomalous velocity producing a transverse force
even in the absence of the Lorentz force~\cite{niu}. This leads to
the anomalous Hall effect (AHE) for the case 
without time-reversal symmetry
~\cite{niu,niu2,juig,hald,kl},
and the SHE for the case 
with time-reversal symmetry~\cite{dya}. 
On the other hand, it was shown by several authors that
the anomalous velocity 
is also generated by spin textures accompanying
spin chirality, and the AHE occurs 
in itinerant electron systems
coupled with localized spins, the orientation of which 
possesses spin chirality order~\cite{ye}.
In this letter, we propose a possible analogue of this phenomenon for 
localized spin systems with no charge degrees of freedom.
We demonstrate that, in analogy with metallic systems, 
the topological Berry-phase effect associated with spin chirality 
raises
the dissipationless spin Hall current induced by
a longitudinal magnetic field gradient:
i.e. Hall effect of spin waves.

Spin current transport associated with spin chirality
was also studied before for one-dimensional magnets;
the Berry phase due to spin textures induces dissipationless
spin currents~\cite{loss,meier},
which parallel persistent charge currents in metal rings threaded by 
a magnetic flux~\cite{beut}.
However, the existence of a transverse force due to the topological Berry phase
 has not been noticed so far except in
a pioneering work done by Haldane and Arovas\cite{halaro},
in which the Hall effect for spin currents in quantum spin systems 
was investigated for the case of two-dimensional (2D)
chiral spin liquids with spin excitation gap,
with emphasis on the quantum Hall effect
characterized by the nonzero Chern number.
However, unfortunately, up to now,
there has been no experimental evidence for the realization
of 2D spin liquid states with spin gap, though
some very recent experiments suggest a possible existence of them
in frustrated magnets~\cite{frust}.
Here, we demonstrate that such exotic transport phenomena
are possible even in more conventional magnets; 
spin Hall currents carried by conventional spin wave
excitations exist in magnetically ordered states~\cite{bubble}.
We will, first, develop a general argument 
for the SHE in spin systems
employing a semiclassical
approach to spin dynamics,
which is similar to a topological description for
the Hall effect in electron systems
developed by Sundaram and Niu~\cite{niu}.
Then, we will present 
an example of a microscopic model in which 
the Hall effect of spin waves occur.

{\it Semiclassical spin dynamics with Berry-phase effects ---}
Here, we consider a semiclassical theory for spin dynamics which is 
in analogy with Sundaram-Niu's topological theory 
for the Hall effect 
of Bloch electrons~\cite{niu}.
In Sundaram-Niu's approach, the dynamics of canonically conjugate
variables $\mbox{\boldmath $x$}$ and $\mbox{\boldmath $k$}$ are described
by semiclassical equations of motion
which include topological Berry-phase effects;
i.e. $\dot{\mbox{\boldmath $x$}}=\nabla_k\varepsilon_k
+\dot{\mbox{\boldmath $k$}}\times \mbox{\boldmath $\Omega$}$,
where $\mbox{\boldmath $\Omega$}$ is the Berry curvature.
The second term is the origin of the intrinsic anomalous Hall effect.
The purpose here is to establish an analogous topological description
for spin transport in localized spin systems. 
We consider spin systems defined on 
a two-dimensional lattice with
the Hamiltonian $\mathcal{H}$ which preserves the total spin projection
$\sum_iS^z_i$.
We assume that spins on nearest neighbor sites $i$ and $j$ 
interacts with each other, and
spin operators $S^{\mu}_i$ with $\mu=x,y,z$
appear in the Hamiltonian in the combinations,
$S^{z}_iS^z_j$,
$S^{+}_iS^{-}_j$, and $S^{-}_iS^{+}_j$. 
We also include the Zeeman term with a spatially varying magnetic field
$\mathcal{H}_{\rm Z}=-g\mu_{\rm B}\sum S^z_iB^z_i$ with
$\nabla B^z \neq 0$,
which plays a role similar to
an electric field in electron systems~\cite{halaro}.
To identify the canonically conjugate variables associated
with spin transport,
we use
the Haldane's representation for spin operators~\cite{hald2},
$S^{+}_i=\sqrt{S+S^z_i}e^{i\hat{\phi}_i}\sqrt{S-S^z_i}$, 
where $S^z_i$ and the phase operator $\hat{\phi}_i$ 
are canonically conjugate variables
satisfying the commutation relation
$[\hat{\phi}_i,S_j^z]=i\delta_{ij}$.
In terms of $S^z$ and $\hat{\phi}$, we define
a new set of canonically conjugate variables,
$\hat{\chi}^{\mu}_{i}=\hat{\phi}_i-\hat{\phi}_{i+\hat{\mu}}$ and
$\hat{\mathcal{T}}_i^{\mu}=-\sum_{\ell_{\mu}=1}^i S^z_{\ell}$,
where $\hat{\mu}$ is a basic lattice vector along the $\mu$-direction
with $\mu=x$ or $y$,
and the sum over the lattice site $\ell$ appeared in the definition of
$\hat{\mathcal{T}}^{\mu}_i$
is taken only for the $\mu$-direction.
These operators satisfy the commutation relation,
$[\hat{\mathcal{T}}^{\mu}_i,\hat{\chi}^{\mu}_j]=i\delta_{ij}$.
The operators $\hat{\chi}^{\mu}$ and $\hat{\mathcal{T}}^{\nu}$
with $\mu\neq \nu$ do not commute with each other.
However, this non-commutativity is not important for the following argument.
The operators $\hat{\mathcal{T}}^{\mu}$ and $\hat{\chi}^{\mu}_i$ are,
respectively, analogous to
an electric dipole operator $e\mbox{\boldmath $x$}$ 
and momentum $\mbox{\boldmath $k$}$ for electron systems. 
In terms of the spin-wave language, $\hat{\phi}$ is the U(1) phase for
the spin wave boson, and thus $\hat{\chi}$, 
i.e. the gradient of $\hat{\phi}$, is
nothing but the U(1) gauge field, i.e. momentum. 
The time-derivative of $\hat{\mathcal{T}}^{\mu}_i$ gives
a spin current  $J^{\mu}_i\equiv d \hat{\mathcal{T}}_i^{\mu}/dt$
which satisfies the continuity equation
$\dot S^z_i+\sum_{\hat{\mu}}[J^{\mu}_{i+\hat{\mu}}-J^{\mu}_i]=0$.

We consider the case that the low energy states are dominated by
gapless magnon excitations from the ground state, and 
the semiclassical approximation for
magnons is applicable, which implies that the expectation values
$\chi^{\mu}_i=\langle u|\hat{\chi}^{\mu}_i|u\rangle$ 
and $\mathcal{T}^{\mu}_i=\langle u|\hat{\mathcal{T}}^{\mu}_i|u\rangle$
with $|u\rangle$ the eigen states of 
$\mathcal{H}_{\rm tot}=\mathcal{H}+\mathcal{H}_{\rm Z}$
can be regarded as continuous variables.
We also assume that different sets of $\{\chi_i^{\mu},\mathcal{T}_i^{\mu}\}$
correspond to different eigen states,
which is valid within the conventional spin wave approximation.
Then, we can introduce a state vector
which is a linear combination of the eigen states,
and a solution for the time-dependent Schr\"odinger equation 
for $\mathcal{H}_{\rm tot}$:
$|\Psi (\mathcal{T}^{\mu}) \rangle=\int d\chi 
a_{\chi}
|u(\chi_i^{\mu},\mathcal{T}_i^{\mu})\rangle $,
where $d\chi=\prod_id\chi_i^{\mu}$
and the normalization condition $\int d\chi |a_{\chi}|^2=1$ is imposed.
The state vector $|\Psi\rangle$ plays a role similar to the wave-packet
function for electron systems~\cite{niu}.
We consider the dynamics of the ``wave-packet'', i.e. the equations of
motion for
$\langle \Psi|\hat{\chi}^{\mu}|\Psi\rangle$ and 
$\langle\Psi |\hat{\mathcal{T}}^{\mu}|\Psi\rangle$, under a spatially slowly
varying external field which yields adiabatic changes of the states
and the Berry-phase effect.
We also introduce a new state $|\tilde{u}\rangle=\hat{U}|u\rangle$ with
$\hat{U}=\exp(-i\sum_i\chi_i^\mu\hat{\mathcal{T}}^\mu_i)
=\exp(i\sum_i\phi_iS^z_i)$.
The operation of $\hat{U}$ amounts to the gauge transformation
in the spin space which
rotates the spin axis at the $i$-th site around the $z$-axis
by an angle $\phi_i$.  
Then, we obtain,
\begin{eqnarray}
\langle\Psi|\hat{\mathcal{T}}^{\mu}_i|\Psi\rangle
=\int d\chi  a^{*}_{\chi}
i\frac{\partial a_{\chi}}{\partial \chi_i^{\mu}}
+\int d\chi d\chi' a_{\chi'}^{*}a_{\chi}
\langle\tilde{u}'|
i\frac{\partial }{\partial \chi_i^{\mu}}
|\tilde{u}\rangle,
\label{tav}
\end{eqnarray}
with $|\tilde{u}'\rangle=|\tilde{u}({\chi_i^{\mu}}',\mathcal{T}^{\mu}_i)\rangle$.
Here we have assumed the orthogonality relation 
$\langle u' | u\rangle=\delta({\chi_i^{\mu}}'-\chi_i^{\mu})$.
Eq.(\ref{tav}) implies that the variable $\hat{\mathcal{T}}^{\mu}$
acts on the function $a_{\chi}$ as an operator defined by
\begin{eqnarray}
\{\hat{\mathcal{T}}^{\mu}_i\}_{\chi\chi'}
=i\frac{\partial}{\partial \chi_i^{\mu}}\delta(\chi-\chi')
+i\langle\tilde{u}'|
\frac{\partial }{\partial \chi_i^{\mu}}|\tilde{u}\rangle.
\label{topera}
\end{eqnarray}
This is regarded as the $\chi$-representation 
for the operator $\hat{\mathcal{T}}^{\mu}$,
and analogous to the crystal momentum representation of the spatial
coordinate $\mbox{\boldmath $x$}$~\cite{blount}.
The second term of (\ref{topera}) is the Berry connection.
In the case with multiple magnon modes, Eq.(\ref{topera}) has
the intra-mode components $\{\hat{\mathcal{T}^{\mu}}\}_{n\chi n\chi'}$
as well as the inter-mode components $\{\hat{\mathcal{T}^{\mu}}\}_{n\chi n'\chi'}$
where $n$ is the index for the $n$-th magnon mode.
The intra-mode components
satisfy the commutation
relation,
\begin{equation}
[\hat{\mathcal{T}}^{\mu},\hat{\mathcal{T}}^{\nu}]
=i\varepsilon_{\lambda\mu\nu}(\mbox{\boldmath $\Omega$}_{\chi\chi})_{\lambda},
\label{tcom}
\end{equation}
where $\mbox{\boldmath $\Omega$}_{\chi\chi}$ is the Berry curvature given by,
\begin{eqnarray}
(\mbox{\boldmath $\Omega$}_{\chi\chi})_{\mu}
=i\frac{\epsilon_{\mu\nu\lambda}}{2}
\left(\left\langle\frac{\partial \tilde{u}}{\partial \chi^{\nu}}
\biggl\vert\frac{\partial \tilde{u}}{\partial \chi^{\lambda}}
\right\rangle
-\left\langle\frac{\partial \tilde{u}}{\partial \chi^{\lambda}}
\biggl\vert\frac{\partial \tilde{u}}{\partial \chi^{\nu}}
\right\rangle\right).
\label{berryc}
\end{eqnarray}
The commutation relation (\ref{tcom}) is a key for topological spin dynamics,
and is in analogy with the commutation relation of $\mbox{\boldmath $x$}$, 
$[x^{\mu},x^{\nu}]=i\varepsilon_{\mu\nu\lambda}\Omega_{\lambda}$,
which is the origin of the intrinsic AHE
for Bloch electrons~\cite{blount}.
The spin dynamics and the spin current are derived 
from the Heisenberg's equation of motion for $\hat{\mathcal{T}}^{\mu}$.
The Zeeman term is rewritten as
$\mathcal{H}_{\rm Z}=-g\mu_{\rm B}\sum_{i}S^z_iB^z_i=
g\mu_{\rm B}\sum_i\hat{\mathcal{T}}_i^{\mu}\nabla_{\mu}B^{z}$
which is analogous to the coupling between an electric dipole 
and an electric field. 
Then, from (\ref{tcom}),
the equations of motion for the expectation values of
$\hat{\mathcal{T}}^{\mu}$ and $\hat{\chi}^{\mu}$
with respect to $\mathcal{H}+\mathcal{H}_{\rm Z}$
are given by,
\begin{eqnarray}
\dot{\mathcal{T}}^{\mu}
=\frac{\partial \mathcal{E}}{\partial \chi^{\mu}}
+g\mu_{\rm B}(\nabla B^z\times\mbox{\boldmath $\Omega$}_{\chi\chi})_{\mu},
\label{eom1}
\end{eqnarray}
\begin{eqnarray}
\dot{\chi}^{\mu}=
g\mu_{\rm B} \nabla_{\mu}B^z,
\label{eom2}
\end{eqnarray}
where $\mathcal{E}=\langle\psi\vert \mathcal{H}\vert\psi\rangle$. 
These equations are, indeed, spin-analogues of Sundaram-Niu's equations for
topological transport of Bloch electrons.
The second term of (\ref{eom1}) is the important Berry-phase effect.
Eq.(\ref{eom2}) describes
the precession motion due to the external magnetic field parallel to 
the $z$-axis, which was obtained previously by Haldane and Arovas~\cite{halaro}.
It is apparent from (\ref{eom1}) and (\ref{eom2})
that the magnetic field gradient $\nabla B^z$ gives rise to
the dissipationless spin Hall current flowing perpendicular to 
the direction of the field-gradient provided that
the Berry curvature (\ref{berryc}) is nonzero.
Note that this Hall effect is more analogous to the AHE for charge currents
rather than the SHE in electron systems,
because time-reversal symmetry is broken in our systems, 
and the spin Hall current is carried 
by the anomalous velocity. 

We, now, discuss the condition for the nonzero
Berry curvature (\ref{berryc}).
In the case of collinear order realized in magnets with
spin rotational symmetry, 
the axis of spontaneous magnetization is parallel to the applied field
along the $z$-direction.
Then, an infinitesimal rotation around
the $z$-axis which corresponds to 
the derivative $\partial |u\rangle /\partial \chi^{\mu}$
does not change the state, and thus,
the Berry curvature vanishes,
$\mbox{\boldmath $\Omega$}_{\chi\chi}=0$.
The existence of non-collinear order (or spin chirality) 
which is generally expected
for frustrated magnets is necessary
for the nonzero $\mbox{\boldmath $\Omega$}_{\chi\chi}$.
However, this is by no means a sufficient condition.
As will be shown later, for the realization of the dissipationless 
SHE,
it is required to introduce parity-breaking terms explicitly into
the Hamiltonian.

{\it Quantum spin model for anomalous spin Hall effect ---}
To demonstrate the existence of the Hall effect of spin waves 
in quantum spin systems,
we present a microscopic model which exhibit the topological transport property
predicted above.
We consider a model of antiferromagnets on a bilayer triangular lattice
with bravais lattice vectors $\mbox{\boldmath $b$}_1=(1,0)$,
$\mbox{\boldmath $b$}_2=(-1/2,\sqrt{3}/2)$. The Hamiltonian is
$\mathcal{H}=\mathcal{H}_1+\mathcal{H}_2+\mathcal{H}_{12}$,
\begin{eqnarray}
&&\mathcal{H}_{n} =J_n\sum_{(i,j)}[S^x_{n,i}S^x_{n,j}+S^y_{n,i}S^y_{n,j}
+\Delta_n S^z_{n,i}S^z_{n,j}]-h_z\sum_iS^z_{n,i}, \nonumber \\
&&\mathcal{H}_{12}=K\sum_{i,\mbox{\boldmath $a$}_{m}}
[S^x_{1,i}S^x_{2,i+\mbox{\boldmath $a$}_{m}}+S^y_{1,i}S^y_{2,i+\mbox{\boldmath $a$}_{m}}],\nonumber
\end{eqnarray}
where
$\mathcal{H}_n$ ($n=1,2$) is the Hamiltonian of a triangular antiferromagnet
on each layer with an in-plane anisotropy $0\leq \Delta_n<1$. 
The sum $\sum_{(i,j)}$ is over pairs of 
the nearest neighbor sites $i,j$. 
$\mathcal{H}_{12}$ is an interaction between two layers,
where the lattice vectors $\mbox{\boldmath $a$}_{m}$ ($m=1,2,3$) are defined by 
$\mbox{\boldmath $a$}_1=\mbox{\boldmath $b$}_1$, 
$\mbox{\boldmath $a$}_2=\mbox{\boldmath $b$}_2$,
and $\mbox{\boldmath $a$}_3=-\mbox{\boldmath $b$}_1-\mbox{\boldmath $b$}_2$. 
The important feature of this model is that the coupling term $\mathcal{H}_{12}$
explicitly breaks inversion symmetry.
As will be shown below, 
this parity-breaking term raises the anomalous velocity of spin waves
associated with the topological Berry phase,
which leads to the Hall effect of spin waves.
In the following, we assume $K\ll J_n$ and 
treat $\mathcal{H}_{12}$ as a perturbation for simplicity.
For $K=0$, 
the ground state of $\mathcal{H}$ with $h_z=0$
is the 120$^{\circ}$ order with all spins aligned parallel to the $xy$-plane.
For small but finite $h_z$, 
all spins are tilted by an angle $\varphi_n$
toward the $z$-direction. The mean field analysis gives
$\sin\varphi_n=h_z/[3J_nS_n(2\Delta_n +1)]$ with $S_n$ the size of spins.
We consider spin transport in this $\varphi$-tilted ordered state, which
possesses a nonzero scalar chirality order
$\chi_{S}\equiv\mbox{\boldmath $S$}_i\cdot(\mbox{\boldmath $S$}_j
\times \mbox{\boldmath $S$}_k)=
\frac{3\sqrt{3}}{2}S_n^3\cos^2\varphi_n\sin\varphi_n$.
To deal with low-energy spin excitations,
we employ a standard spin wave theory within
the Gaussian approximation~\cite{miyake,chu}.
We also postulate that infinite numbers of the bilayer systems are
weakly coupled via a inter-bilayer interaction 
to stabilize the magnetic order 
at finite temperatures.


In this system, 
only the in-plane spin wave mode is gapless.
This in-plane mode carries a spin current with the magnetization parallel to
the $z$-axis.  
The spin current $J_z(i,j)$ on the bond $(ij)$, which 
satisfies the continuity equation, 
is obtained from $\dot{S}^z_i=i[\mathcal{H},S^z_i]=
-\sum_m[J_z(i,i+\mbox{\boldmath $a$}_m)
-J_z(i-\mbox{\boldmath $a$}_m,i)]$~\cite{halaro}.
The spin current consists of three parts, i.e.
$J_{z}(i,j)=J^{11}_{z}(i,j)+J^{22}_{z}(i,j)+J^{12}_z(i,j)$.
Here, the first two terms are intra-plane spin currents: 
$J^{nn}_{z}(i,j)=
J_n(\mbox{\boldmath $S$}_{n,i}\times\mbox{\boldmath $S$}_{n,j} )_z$.
The third term $J^{12}_{z}(i,j)=K
(S^x_{1,i}S^y_{2,j}-S^y_{1,i}S^x_{2,j})$, which stems from the inter-layer coupling
$\mathcal{H}_{12}$, gives rise to the anomalous spin Hall current. 
The $x$- and $y$-components of the spin current are, respectively,
$J_{z,x}(i)=J_z(i,i+\mbox{\boldmath $a$}_1)-\frac{1}{2}[J_z(i+\mbox{\boldmath $a$}_2,i)
+J_z(i+\mbox{\boldmath $a$}_3,i)]$, and
$J_{z,y}(i)=\frac{\sqrt{3}}{2}[J_z(i,i+\mbox{\boldmath $a$}_2)
-J_z(i+\mbox{\boldmath $a$}_3,i)]$.

\begin{figure}
\includegraphics[width=8.5cm]{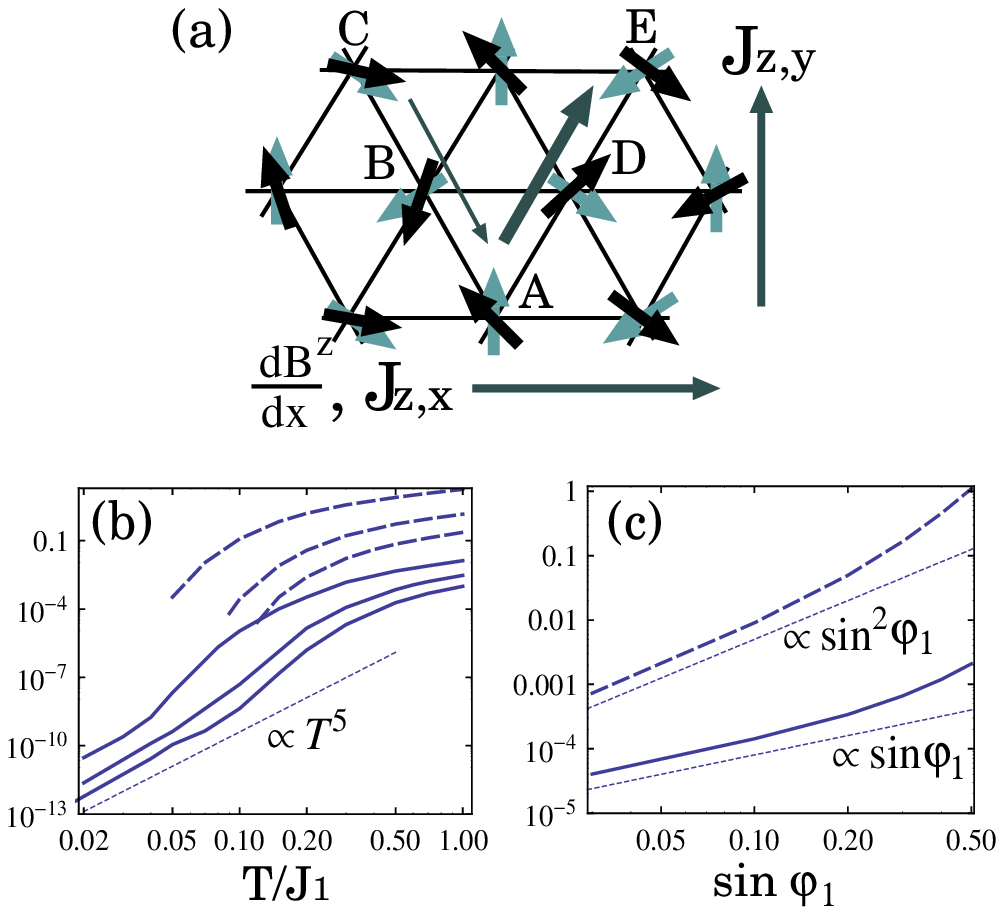}
\caption{(a) Schematic view of the origin of the Hall effect of spin waves. 
Short arrows represent spin structures.
Long arrows represent spin current flows.
See the text. 
(b) Examples of plots of 
$\sigma^{\rm SHE}_{xy}/K^2$ (solid line) and 
$\sigma^{\rm D}_{xy}/\tau J_1^3$ (broken line) versus $T/J_1$ 
for several values of $\sin \varphi_1$. 
$\Delta_1=\Delta_2=0.8$, $J_1=1.0$, $J_2=1.5$. 
$\sin\varphi_1=0.3, ~0.5, ~0.8$ (from bottom to top).
(c) Plots of $\sigma^{\rm SHE}_{xy}/K^2$ (solid line)
and $\sigma^{\rm D}_{xy}/\tau J_1^3$ (broken line) versus 
$\sin\varphi_1$
for $\Delta_1=\Delta_2=0.8$, $J_1=1.0$, $J_2=1.5$,
and $T/J_1=0.8$.
}
\end{figure}

When there is a magnetic field gradient, e.g.
$\partial B^z/\partial x\neq 0$, the spin Hall current
$J_y=\sigma^{\rm SHE}_{yx}\partial B^z/\partial x$ appears. 
The origin of the spin Hall effect in this model is schematically understood as follows.
As shown in FIG. 1(a), in the $\varphi$-tilted 120$^{\circ}$ ordered state, 
(the spin structure is represented by gray arrows), 
the $y$-component of
the spin current carried by the in-plane spin wave mode flowing along the path 
A$\rightarrow$B$\rightarrow$C and 
that flowing along the path A$\rightarrow$D$\rightarrow$E 
cancel with each other, 
and there is
no net spin current along the $y$-direction.
When the field gradient parallel to the $x$-axis, 
$\partial_x B^z$, is applied, the spin structure
is changed to that represented by the black arrows.
Then, the spins along these two paths 
feel a different torque, 
because of the difference of the field strength among
the sites B, C, D, and E. 
As a result, the spin currents along the path A$\rightarrow$B$\rightarrow$C 
and that along the path A$\rightarrow$D$\rightarrow$E are unbalanced, and
the spin Hall current along the $y$-axis, $J_{z,y}$, appears.
Using the Kubo formula, 
we calculate, up to the lowest order in $K$,
the Hall conductivity for the spin current 
$J^{12}_{z,\mu}=\sum_iJ^{12}_{z,\mu}(i)$ ($\mu=x,y$),  
\begin{eqnarray}
\sigma^{\rm SHE}_{xy}=\lim_{\omega\rightarrow 0}\frac{i}{\omega}
\int^{\infty}_0dt\langle[J^{12}_{z,x}(t),J^{12}_{z,y}(0)]\rangle e^{i\omega t}
\label{sig}
\end{eqnarray}
with 
$\langle ...\rangle$ the average
with respect to $\mathcal{H}_1+\mathcal{H}_2$.
We found that 
$\sigma^{\rm SHE}_{xy}$ is nonzero when two layers are not equivalent; i.e.
$J_1S_1\neq J_2S_2$ or $\Delta_1\neq\Delta_2$. 
This condition is analogous to the condition of spin imbalance 
for the anomalous Hall effect in electron systems~\cite{kl}. 
This implies that the Hall effect of spin waves is 
analogous to the AHE for charge currents 
in electron systems rather than the SHE in
electron systems with time reversal symmetry.
Actually, in our systems, time reversal symmetry 
is broken by both magnetic order and magnetic fields.
Also it is noted that the role played by spin degrees of freedom
for the AHE in electron systems is
played not by spins but by two-layer degrees of freedom in our spin system. 
In FIG. 1(b), we show an example of temperature dependence of
$\sigma^{\rm SHE}_{xy}$ calculated numerically.
At low $T$, $\sigma_{xy}^{\rm SHE} \propto T^5.$
The Hall conductivity (\ref{sig}) does not depend on 
the relaxation time of magnons $\tau$. 
The $T$-dependence of $\sigma^{\rm SHE}_{xy}$ is merely due to
thermally excited carriers, i.e.
the Bose distribution function. 
This dissipationless Hall effect 
is raised by the Berry curvature
(\ref{berryc})
associated with the scalar chirality, as predicted from the semiclassical 
analysis, Eq.(\ref{eom1}).

Note that in our model, 
in addition to the dissipationless Hall effect, 
the dissipative Hall effect, which depends on the relaxation time $\tau$, 
is also possible even when $K=0$.
The dissipative Hall effect is analogous to
the extrinsic AHE of electron systems.
However, the analogy is not complete.
In contrast to the extrinsic AHE, the origin of the transverse force 
on spin waves is not asymmetric scattering, but the existence of
the spin scalar chirality.
The dissipative Hall effect is more deeply related to spin currents induced by
magnetic fields in one-dimensional magnets with noncoplanar 
spin texture~\cite{loss}.
In two or three dimensional systems, for a particular spin texture, 
it is possible that a field gradient induces transverse spin current
as illustrated in FIG.1(a).
This effect is dissipative in the sense that the Hall conductivity generally
depends on the relaxation time of magnons. 
The dissipative spin Hall conductivity $\sigma^{\rm D}_{xy}$ 
is derived from Eq.(\ref{sig})
with $J^{12}$ replaced with $J^{11}$ or $J^{22}$.
The total Hall conductivity is, then, given by 
$\sigma^{\rm SHE}_{xy}+\sigma^{\rm D}_{xy}$.
We assume that $\tau$ is governed by impurity scattering, 
and independent of temperatures.
Then, the numerically obtained 
$\sigma^{\rm D}_{xy}$ is order of magnitudes larger than
$\sigma^{\rm SHE}_{xy}$ (FIG. 1(b)).
This makes it difficult to detect the dissipationless effect.
Nevertheless, the dissipationless and dissipative contributions to the 
Hall effects are distinguishable
from the different dependence on $\sin\varphi$, i.e. $\chi_S$.
We can show analytically that up to the lowest order in $\chi_S$,
$\sigma^{\rm SHE}_{xy}\propto \chi_S$ and 
$\sigma^{\rm D}_{xy}\propto \chi_S^2$ holds.
At high temperatures $T\sim J_1$,
these relations are retained up to $\sin\varphi <0.3$ (FIG. 1(c)).
Thus, the different dependence on $\chi_S$ 
clearly 
distinguishes the dissipationless Hall effect from
the dissipative one.



{\it Concluding remarks---}
Although the model presented above is a toy model which does not describe
actual magnetic systems known so far, its realization in real materials
is feasible.
Furthermore, 
it establishes the concept of the Hall effect of spin wave excitations
caused by both the dissipationless and dissipative mechanisms.
Our results open a possibility of exploring the SHE
in localized spin systems with no charge degrees of freedom.
The current study can be extended to some other transport
properties such as 
thermal Hall currents carried by spin waves.
This is also related to a spin analogue of the Nernst effect.
This issue will be addressed in the near future.

Finally, it is noted that
the nonzero Berry curvature in our spin system strongly 
suggests the existence of topological order, 
as in the case of the AHE~\cite{niu,hald}.
It is curious to examine this scenario, 
when the spin wave excitations acquire a gap due to
a spin anisotropy.
However, for this purpose, we need to evaluate the Chern number
dealing with the inter-layer interaction $\mathcal{H}_{12}$
non-perturbatively.
Although we do not pursue this direction here, this issue
deserves further precise investigations.

In summary, we have demonstrated that in a certain class of
frustrated magnets, 
the topological Berry-phase effect associated with spin chirality 
gives rise to the Hall effect of spin waves, in analogy with 
the AHE in electron systems.

We thank Qian Niu for valuable discussions, which helped us to clarify
 the dissipationless Hall effect in our microscopic model.
We also thank one of the referees who suggests a possible relation with
topological order.
This work is supported by the Grant-in-Aids for
Scientific Research from MEXT of Japan
(Grant No.18540347, Grant No.19014009, Grant No.19052003).

\end{document}